# Comment: Monitoring Networked Applications With Incremental Quantile Estimation

Lorraine Denby, James M. Landwehr and Jean Meloche

Monitoring networked applications is indeed a challenge, particularly so for large networks. There are many types of specific networked applications, and a range of statistical and computational issues comes up in different problems. Chambers et al. discuss the constraints and their solution for a specific business application software problem they faced. While we would like to know more about the real-world nature of their application and how the users' needs and the application's technology drive the constraints they faced, the statistical approach they developed and its results are impressive. Problems that we have faced have somewhat different technological and statistical needs, however. We would like to take this opportunity to describe some general issues and approaches that we believe are useful for networked application monitoring problems.

Given the quality of today's networks and applications with relatively little downtime and stable behavior, the monitoring process often amounts to filtering through large amounts of irrelevant data and looking for the unusual. The monitoring system can be regarded as a compression engine that takes the large amounts of irrelevant data and distills them into the few elements of information that do matter.

The raw data that come into the compression engine originate at multiple agents in the network, often but not exclusively at the endpoint devices of the network where the application clients run. Additional data can also be available from the network elements (e.g., routers and links) that actually handle the traffic. The ultimate destination is often but not necessarily a centralized system where it will be possible to launch corrective action as need be. The frequency with which the raw data arise, the number of agents, the speed at which unusual events are reported, and the size of the reported information are all examples of technical parameters of the monitoring system that are at the root of the challenge.

In every instance of this problem, the technical parameters are constrained by business requirements. The business requirements aim to specify characteristics of the monitoring system such as the speed with which it will notice developing problems, the frequency of false alarms, and the network overhead (amount of network traffic dedicated to the monitoring itself) of the system. The network overhead is an especially difficult constraint to handle in general because it relates to specific characteristics of the network on which the distributed application is running and it requires cost considerations. For example, a large overhead on an optical segment of the network may be of no consequence in comparison to a moderate one on a low-bandwidth link.

For some combinations of the technical parameters and business constraints, it may be possible to simply send all of the raw data to a central server for analysis. The paper that we are discussing starts from the premise that this is not the case. The problem then faced is to design something like a distributed compression machine. Chambers et al. propose a model in which the agents perform part of the compression and send partly summarized data to a central server where the aggregation of the various summaries takes place. First, the agents fill a data buffer $D$ of size $N$; second, when $D$ is full a quantile buffer $Q$ is updated and $D$ is flushed; third, periodically or upon request, $Q$ is sent to the server for aggregation.

The problem is related to that of distributed source coding that has been studied in computer science.

*Lorraine Denby, James M. Landwehr and Jean Meloche are with the Data Analysis Research Department, Avaya Labs, Basking Ridge, New Jersey 07920, USA e-mail: ld, jml, jmeloche@research.avayalabs.com.*

In that context, the agents transmit coded signals to a server where the decoding takes place. The primary goal is to find coding methods that result in a tolerable distortion.

As a first specific and important example, we consider VoIP (Voice over IP). VoIP typically involves a pair of IP phones that send an RTP (real-time protocol) stream to each other. The RTP stream is a sequence of numbered UDP (user datagram protocol) packets. At the sending phone, the packets are sent with strict regularity (every 20 milliseconds, e.g.) and each packet contains a sample of voice (20 milliseconds of voice in the example). At the receiving phone, the packets do not arrive so regularly because of uncontrollable events in the network. In addition to the RTP stream, the IP phones periodically (every 5 seconds usually) exchange RTCP (real-time control protocol) packets which hold summary statistics that report what are considered to be the relevant aspects of the RTP stream performance, namely, the observed loss, delay and jitter (irregularity) in the packet arrival times. The Avaya IP phones can be configured to send a copy of the RTCP packets to a central repository. The information collected in this fashion is invaluable for diagnostic purposes and enables the server to take decisions regarding call routing. The timeliness of the decision is critical to the high quality of VoIP. The network footprint, however, can easily become prohibitive when there are tens of thousands of endpoint devices sending RTCP packets to a central repository every 5 seconds. With colleagues, we have explored aspects of assessing networks for VoIP QoS (quality of service) in Bearden et al. (2002a, 2002b) and proposed methods for scalability in Karacali, Denby and Meloche (2004). An approach for distributed monitoring and analysis was described in Adhikari et al. (2006).

A similar example arises in the context of the monitoring of a service-level agreement (SLA) for a number of locations where the monitoring agents are located. Agreements are often expressed in terms of a distribution for end-to-end packet transit time over stipulated periods of time. The client of the agreement will want to monitor the end-to-end packet transit times and document violations of the agreement. In such a situation it may be desirable to detect the violations quickly in order to take corrective action before too much damage is done.

A third example, where the underlying data are on a completely different time scale, is that of a distributed contact center. An enterprise's contact center (e.g., to handle customer service) can have agents (for this example, human beings) located at many sites but with their work assignments handled through an application routing the voice calls among the sites and agents. The agents may be distributed geographically among several large physical sites, and/or they may also be distributed among several "workgroups" of people with similar skills and responsibilities regardless of their geographic location. Interesting events here concern lengths of time for agents to handle segments of a customer call, so the relevant time scales are in seconds and minutes, not milliseconds as with VoIP quality of service. Nevertheless, the overall contact center application has two main monitoring needs for this problem: first, to monitor the status of agents and calls so that both customers and agents are treated equitably and the workload is distributed efficiently by the call-routing engine; doing so requires having some central status notions so that real-time routing decisions are made satisfactorily. Second, reporting on the operation of the distributed contact center both real-time and historically requires summarizing and storing information about the customer calls and the agents who handled them so that the results can be summarized by different groupings such as agent, workgroup, geographic location, individual customer over several calls, and customer segments; and all over appropriate time periods.

With the above examples in mind, we propose a statistical approach different from the technology developed by Chambers et al. for their problem. Let the agents maintain a buffer of the last $N$ time-value pairs and send to the server a subset of $M$ of those pairs whenever the distribution of the $N$ most recent values justifies a transmission. For example, the agents could send a sample when the distribution of recent values differs significantly from the distribution of values that was last reported by the agent, using for example a Mann–Whitney $U$ test. Alternatively, in an effort to minimize the network overhead, the agent could attack this as a change-point problem and avoid sending all but the recent data since the server is already well informed of the distribution of values that prevailed before the agent noticed the change. The server can integrate the new data and perform analyses for which the aggregation is necessary. In general, the problem amounts to finding a function $f$ of $N$ variables that indicates which of the buffered values should be sent to the server.

In a steady state, $f$ would return the empty set. The function $f$ could be selected so that the server can estimate various parameters efficiently.

Formally, we have agents $1 \leq i \leq K$ observing $X_{ij}$ at time $t_{ij}$ for $j \geq 1$. If the network overhead could be ignored, we would send all of the $X_{ij}$ to a central server and we could evaluate things such as the CDF $F_t$ of all of the observed values in the last $t$ seconds. In our scheme, the effort to minimize the network overhead amounts to a careful selection of the time-value pairs that are sent to the server and the agents maintain their buffer of values for the purpose of deciding which values are worth sending to the server.

There are several advantages to what we propose. First, while the system is in control, no data are sent from agent to server at all, avoiding the unnecessary network overhead of sending the same old summary. Second, the server gets evidence of a change as soon as possible. In real-time monitoring scenarios in which some automatic or human-initiated action can result from a change in values, this can reduce response time and be an important benefit. Finally, the method proposed by the authors of the paper being discussed assumes that the data stored in the agent buffer are interchangeable, an assumption which is not compatible with the possibility of rapidly developing problems.

Evidently all methods have advantages and disadvantages. A comparison between candidate monitoring systems should include both business requirements of the monitoring system as well as technical capabilities of potential solutions. Ideally one would like to evaluate technical options versus given business requirements, but in reality there is always iteration between the two components. A list of network characteristics that deserve consideration should include the number of agents, speeds at which raw data arise (possibly different speeds at each agent), the network topology, and the types of events that need to be detected or reported on. As the authors point out, aspects such as the memory and CPU requirements on the agents and the server are also important. Statistical aspects for evaluating possible technical solutions surely should include characteristics such as reaction times to detect important events, the false alarm rate, and the network load generated globally and locally by the statistical monitoring technology.

There is a wealth of interesting and important problems in the arena of distributed monitoring of networked applications. Our impression is that, to date, these have been addressed primarily by computer scientists and computer networking engineers and researchers. We believe that this area presents interesting, challenging and fruitful opportunities for statisticians to make contributions. We thank the journal Editors and authors of this paper for helping to bring this topic to the attention of a wide statistical audience.